\documentclass{article}
\usepackage{PRIMEarxiv}
\usepackage[utf8]{inputenc} % allow utf-8 input
\usepackage[T1]{fontenc}    % use 8-bit T1 fonts
\usepackage{hyperref}       % hyperlinks
\usepackage{url}            % simple URL typesetting
\usepackage{booktabs}       % professional-quality tables
\usepackage{amsfonts}       % blackboard math symbols
\usepackage{nicefrac}       % compact symbols for 1/2, etc.
\usepackage{microtype}      % microtypography
\usepackage{lipsum}
\usepackage{multirow}
\usepackage{color}
\usepackage{tikz}
\usepackage{import}
\usepackage{textcomp, gensymb}
\usepackage[autosize]{dot2texi}
\usepackage{tikz}
\usetikzlibrary{shapes,arrows}
\usepackage{caption}
\usepackage{subcaption}
\usepackage{amsmath}
\usepackage{geometry,mathtools}
\usepackage{graphicx}
\usepackage{pstool}
\usepackage{fancyhdr}       % header
\usepackage{graphicx}       % graphics
\graphicspath{{media/}}     % organize your images and other figures under 
\usepackage{algorithm}
\usepackage{algpseudocode}
\usepackage{lineno,hyperref}
\usepackage{amsmath,amsfonts}
\usepackage{url}            % simple URL typesetting
\usepackage{import}
\usepackage{gensymb}
\usepackage{caption}
\usepackage{subcaption}
\usepackage{amssymb}
\usepackage{textcomp}
\usepackage{stmaryrd}
\usepackage{xcolor}
\usepackage{pstool}
\usepackage{stmaryrd}
\usepackage{geometry,mathtools}
\usepackage{graphicx}
\usepackage{pstool}
\usepackage{graphicx}       % graphics
\graphicspath{{media/}}     % organize your

\usepackage{array}
\usepackage{color,soul}

\title{Low-density parity-check codes: tracking non-stationary channel noise using sequential variational Bayesian estimates
}

\author{
  J du Toit, JA du Preez, and R Wolhuter \\
  Electrical and Electronic Engineering \\
  Stellenbosch University \\
  Stellenbosch, South Africa \\
  \texttt{jacowp357@gmail.com, dupreez@sun.ac.za, wolhuter@sun.ac.za} \\
}
\begin{document}
\maketitle

\begin{abstract}
We present a sequential Bayesian learning method for tracking non-stationary signal-to-noise ratios in low-density parity-check (LDPC) codes by way of probabilistic graphical models. We represent the LDPC code as a cluster graph using a general purpose cluster graph construction algorithm called the layered trees running intersection property (LTRIP) algorithm. The channel noise estimator is a global gamma cluster, which we extend to allow for Bayesian tracking of non-stationary noise variation. We evaluate our proposed model on real-world 5G drive-test data. Our results show that our model can track non-stationary channel noise accurately while adding performance benefits to the LDPC code, which outperforms an LDPC code with a fixed stationary knowledge of the actual channel noise.
\end{abstract}

\keywords{LDPC codes, Bayesian sequential learning, variational inference, cluster graphs}

\section{Introduction}
\label{sec:introduction}
In wireless communication systems, channel noise interferes with radio transmissions between a transmitter and receiver. The nature of the noise is assumed to be non-stationary, since it can change over time and vary according to environmental dynamics.

Knowledge of channel noise is useful in a communication system to adapt certain radio parameters, ultimately optimising the reliability and overall quality of communication. These optimisation methods include, inter alia, adaptive modulation and coding (AMC), interference suppression, and rate-matching.

Communication systems such as LTE and 5G rely on predefined reference signals, i.e. pilot signals, which are used to estimate the channel noise. These pilot signals are generated in the physical layer and scheduled periodically via the physical uplink control channel (PUCCH), or aperiodically via the physical uplink shared channel (PUSCH)~\cite{rumney2013lte}. Instantaneous estimates of the channel noise should ideally be acquired at the same rate as the channel changes, which require frequent use of pilot signals. The downside of doing so is an additional communication overhead, since the time-frequency resources could be utilised to carry payload data instead~\cite{massivemimobook}.

Our interest is in finding a way to estimate channel noise without relying solely on scheduled pilot signals. The aim of our work is not to omit the broader channel state information (CSI) system in its entirety, since it also provides information on other channel properties outside the scope of this work. Instead, our focus is on reliably and dynamically estimating the channel noise only, which we see as a step towards substituting certain elements of the CSI system. We propose to combine received signal values from the physical layer with a channel-coding scheme such as a \emph{low-density parity-check} (LDPC) code. By doing so, we extend a standard LDPC code by integrating a channel-noise estimator into the decoding algorithm itself, which can learn the channel noise on-the-fly and simultaneously benefit from the channel-noise estimates.

LDPC codes were first introduced by Gallager in 1962 and rediscovered in 1995 by MacKay~\cite{gallager1962low, davey1998low}. These codes are a family of block codes with good theoretical and practical properties. They have found extensive application in various wireless communication systems, including the 5G New Radio (NR) technology standard, which will be used in our study.

While knowledge of the channel noise is useful for aspects of the communication system discussed before, LDPC decoders also benefit from this information. LDPC codes require knowledge of the channel noise variance when the communication channel is assumed to be an \emph{additive white Gaussian noise} (AWGN) channel~\cite{johnson2006introducing}. Channel noise variance corresponds to the channel \emph{signal-to-noise ratio} (SNR), which we will use interchangeably in this paper to refer to the channel noise. If the noise variance is over- or understated, the performance of the LDPC decoder can suffer~\cite{saeedi2007performance,tan2004signal,mackay2003performance}.

Alternative solutions, including the focus of our work, aim to estimate SNRs in the \emph{forward error correction} (FEC) layer while concurrently decoding received messages. The rationale is that through the use of LDPC codes, the channel uncertainty can be made explicit in the LDPC decoder rather than assuming it to have a fixed ``confident'' a priori value. In this regard, we draw inspiration from a famous quote by Mark Twain: ``\emph{It ain't what you don't know that gets you into trouble. It's what you know for sure that just ain't so.}'' By making the decoder ``aware'' of its uncertainty (i.e. allowing it to learn the channel constraints), it may also improve the error-correcting capability of the LDPC code. We use a Bayesian approach that takes into account a statistical distribution over the SNR, which is also a much better realisation of the channel in terms of its stochastic time-varying nature.

% How can SNRs be estimated in the LDPC decoder?
Iterative \emph{message passing} algorithms used in LDPC codes such as
\emph{bit-flipping} use binary (hard-decision) messages between
\emph{variable} nodes and \emph{check} nodes to update the received
bit values in a \emph{factor} graph (also known as a \emph{Tanner}
graph). Whereas this decoding method is fast, it does not provide
probabilistic information about the received bits, which is required
to estimate an SNR statistic in AWGN channels. The \emph{sum-product}
algorithm uses probabilistic (soft-decision) messages and is
equivalent to the \emph{loopy belief propagation} (LBP) algorithm used
for performing generic inference tasks on \emph{probabilistic
graphical models} (PGMs)~\cite{mackay2003information, Koller2009}.
However, as will become clear in
Section~\ref{sec:ldpc-with-snr-estimation}, modelling SNRs in AWGN
channels results in the use of conditional Gaussian distributions that
during subsequent message passing morphs into mixture distributions --
these have problematic properties for inference that force us to
employ approximation techniques.

% What other solutions exist?
As such, studies in~\cite{wang2011noise,jakubisin2016probabilistic, wu2014expectation,senst2012message,nissila2009adaptive,wang2011noise2,cui2012online} have proposed hybrid methods for jointly modelling the SNRs and the transmitted bit messages. These methods are all based on factor graphs with some graph extension that uses an inference scheme different from LBP for estimating the SNRs. A study in \cite{senst2012message} presents a comparison between a \emph{variational message passing} (VMP) based estimator and an \emph{expectation maximisation} (EM) based estimator for stationary SNRs. The VMP-based estimator demonstrated superior performance over the EM-based estimator and achieved a lower frame error rate, with no significant increase in computational complexity. In \cite{wang2011noise,wang2011noise2,cui2012online} SNRs are assumed stationary within fixed independent sections of the LDPC packet with a focus on finding a low complexity noise estimator.
% What other solutions lack?
These studies do not model the sequential dynamics of channel noise variance, e.g. the possible correlations between inferred instances of SNRs.

% What is our approach and assumptions for estimating the channel noise?
It is reasonable to assume that the same noise energy can influence an entire LDPC packet over the duration of a packet due to the transmission speed at which communication takes place. The basic transmission time interval in LTE systems, for example, is 1 ms~\cite{dahlman20134g}. Furthermore, a succeeding LDPC packet's noise estimate can also depend on the noise induced on previous LDPC packets. We introduce a dependency between LDPC packets that gradually ``forgets'' previous channel noise estimates. This allows an LDPC decoder to track time-varying SNRs as packets arrive at the receiver.

Initially, we used a standard factor graph representation with an additional continuous random variable to model the SNR at a packet level. The larger parity-check factors from the LDPC code created computational challenges, which led us to review other options for representing the problem and doing inference. To address this, we (1) use a more general graphical structure called \emph{cluster} graphs, (2) use a variant of LBP called \emph{loopy belief update} (LBU), (3) create a message order that relaxes repeated marginalisation from large parity-check factors, and (4) introduce a cluster-stopping criterion that turns off uninformative clusters during message passing. Cluster graphs have been shown to outperform factor graphs in some inference tasks in terms of accuracy and convergence properties~\cite{Koller2009,yedidia2005constructing,streicher2017graph}.

\textbf{Our contribution}: We view this problem more generally as a PGM and introduce a \emph{sequential Bayesian learning} technique capable of tracking non-stationary SNRs over time. We represent the PGM as a cluster graph compiled by means of a general purpose algorithm called \emph{layered trees running intersection property} (LTRIP), developed in~\cite{streicher2017graph}. We believe this is the first work that represents LDPC codes as a cluster graph, which may be due to a lack of available algorithms that can construct valid cluster graphs. We demonstrate: (1) tracking of non-stationary channel noise over time, and (2) performance benefits of our approach compared to an LDPC code with stationary knowledge of the channel as well as an LDPC code with perfect knowledge of the channel.

This paper is structured as follows: In Section~\ref{sec:ldpc-with-snr-estimation}, we explain how LDPC codes are represented in a more general PGM framework and extended with the channel noise estimator. Our message passing approach and schedule are also explained. Section~\ref{sec:bayesian-sequential} describes how the PGM is updated sequentially, which allows the model to track non-stationary channel noise. The results are shown in Section~\ref{sec:experimental-investigation} and in Section~\ref{sec:supplementary-investigation} we present a supplementary investigation. Finally, our conclusions and future work are discussed in Section~\ref{sec:conclusion}.

\section{LDPC codes with non-stationary SNR estimation as PGMs}
\label{sec:ldpc-with-snr-estimation}
In this section, we compile a cluster graph of an LDPC code and introduce additional random variables that are used to estimate the channel noise. This is extended further to enable the tracking of non-stationary channel noise. We also discuss our hybrid inference approach and message passing schedule.

\subsection{Channel SNR estimation with LDPC codes}
We assume an AWGN channel model (without fading), which adds zero mean
random Gaussian noise to a transmitted signal. The strength of the
noise depends on the Gaussian precision (the inverse of the
variance). We denote the LDPC code's bit sequence as $b_0,...,b_N$,
where $N$ is the length of the codeword. We use BPSK signal modulation
with unit energy per bit $E_b = 1$ (i.e. a normalised signal). The
channel-noise precision is an unknown quantity for which we assign a
gamma prior distribution -- a Bayesian way of treating unknown
variables. The gamma distribution is the conjugate prior for the
precision of a Gaussian distribution and is part of the exponential
family~\cite[Section 2.3.6]{bishop2006pattern}. With these
assumptions, the observed received signal $x_n$ is modelled using a
conditional Gaussian likelihood function $f(x_n \mid b_n,\mu,\gamma)$
with known means $\mu \in$
\{$\mu_{0}=-\sqrt{E_b}=-1,\mu_{1}=\sqrt{E_b}=1$\} that represent the
modulated binary data (i.e. the two different phases of the carrier
wave). Since the same noise influences both states of $b$ the received
signal's likelihood function is simplified to:
\begin{equation}
f(x_{n} \mid b_{n},\gamma) =
\begin{cases}
\sqrt{\frac{\gamma}{2}}e^{-\frac{\gamma(x_n + 1)^2}{2}} & \text{when } b_n = 0, \\
\sqrt{\frac{\gamma}{2}}e^{-\frac{\gamma(x_n - 1)^2}{2}} & \text{when } b_n = 1.
\end{cases}
\label{eq:conditional-gaussian}
\end{equation}

After multiplying in the prior distributions over $b_n$ and $\gamma$,
the presence of the $b_n$ terms in the joint changes this to a mixture
distribution which does not form part of the exponential family.  This
implies that its marginals towards the $\gamma$ and $b_n$ variables
will not be conjugate towards their prior forms. To rectify this
requires some form of approximation that forces these marginals to the
required conjugate forms. In the work here we do this by using the VMP
approach -- thereby replacing the sufficient statistics of $\gamma$
and $b_n$ random variables with their expected values. This leaves
Equation~\ref{eq:conditional-gaussian} in its conditional form~\cite[Section 4.3]{winn2005variational} --
Section~\ref{sec:vmp} provides the details about this.

The channel noise is assumed independent and identically distributed (i.i.d.) over the length $N$ of an LDPC packet (the codeword length). The transmission duration of a packet is around 1ms~\cite{dahlman20134g}, which we assume provides sufficient samples for estimating the noise, and is a short enough time frame for non-stationarity to not be a problem. The channel-noise precision can be translated to a rate-compensated SNR given by $\textrm{SNR}_{\textrm{dB}} = 10 \log_{10}(\frac{E_b\gamma}{2R})$, where $R$ is the code rate~\cite[Section 11.1]{mackay2003information}, and $\gamma$ is the distribution over the channel-noise precision. Note that we may use the terms precision and SNR interchangeably. The next section presents LDPC codes with SNR estimation using the gamma prior and conditional Gaussian random variables discussed here.

\subsection{Representation of LDPC codes with SNR estimation}
\label{sec:representation}
During the course of this study, we noted that cluster graph representations of LDPC codes are not addressed in the available channel coding literature. Researchers in the channel coding domain may be more familiar with the factor graph (or Tanner graph) representation of LDPC codes. While our study focuses on channel noise estimation, we address the novelty and performance advantages of cluster graph LDPC codes compared to factor graph LDPC codes in a subsequent study~\cite{jacojohancluster}. Nonetheless, in the interest of readability, we offer a brief summary of cluster graphs here without diminishing the importance of our other study.

A cluster graph is an undirected graph consisting of two types of
nodes. A cluster node (ellipse) is a set of random variables and a
sepset (short for ``separation set'') node (square) is a set of random
variables shared between a pair of clusters. In most interesting cases
(which also include LDPC decoding), this graph will not be a tree
structure, but will contain loops. Inference on such a ``loopy''
system requires the so-called \emph{running intersection property}
(RIP)~\cite[Section 11.3.2]{Koller2009}. This specifies that any two
clusters containing a shared variable, must be connected via a
\emph{unique} sequence of sepsets all containing that particular
variable, i.e. no loops are allowed \emph{for any particular
variable}.

Our study uses a general purpose cluster graph construction algorithm that produces a cluster graph of the LDPC code. This algorithm is termed the \emph{layered trees running intersection property} (LTRIP) algorithm developed in~\cite{streicher2017graph}. The LTRIP algorithm proceeds by processing layers, with each layer dedicated to an individual variable. For each variable, it determines an optimal tree structure over all clusters. The final sepsets are formed by merging the sepsets between pairs of clusters across all layers. The resulting cluster graph satisfies the RIP. Compared to factor graphs, cluster graphs offer the benefits of more precise inference and faster convergence. We refer the reader to~\cite{streicher2017graph} for more detail regarding the LTRIP algorithm and~\cite{jacojohancluster} for a comparison study between cluster graph and factor graph LDPC codes.

\begin{figure}[h]
\centering
\includegraphics[width=\textwidth]{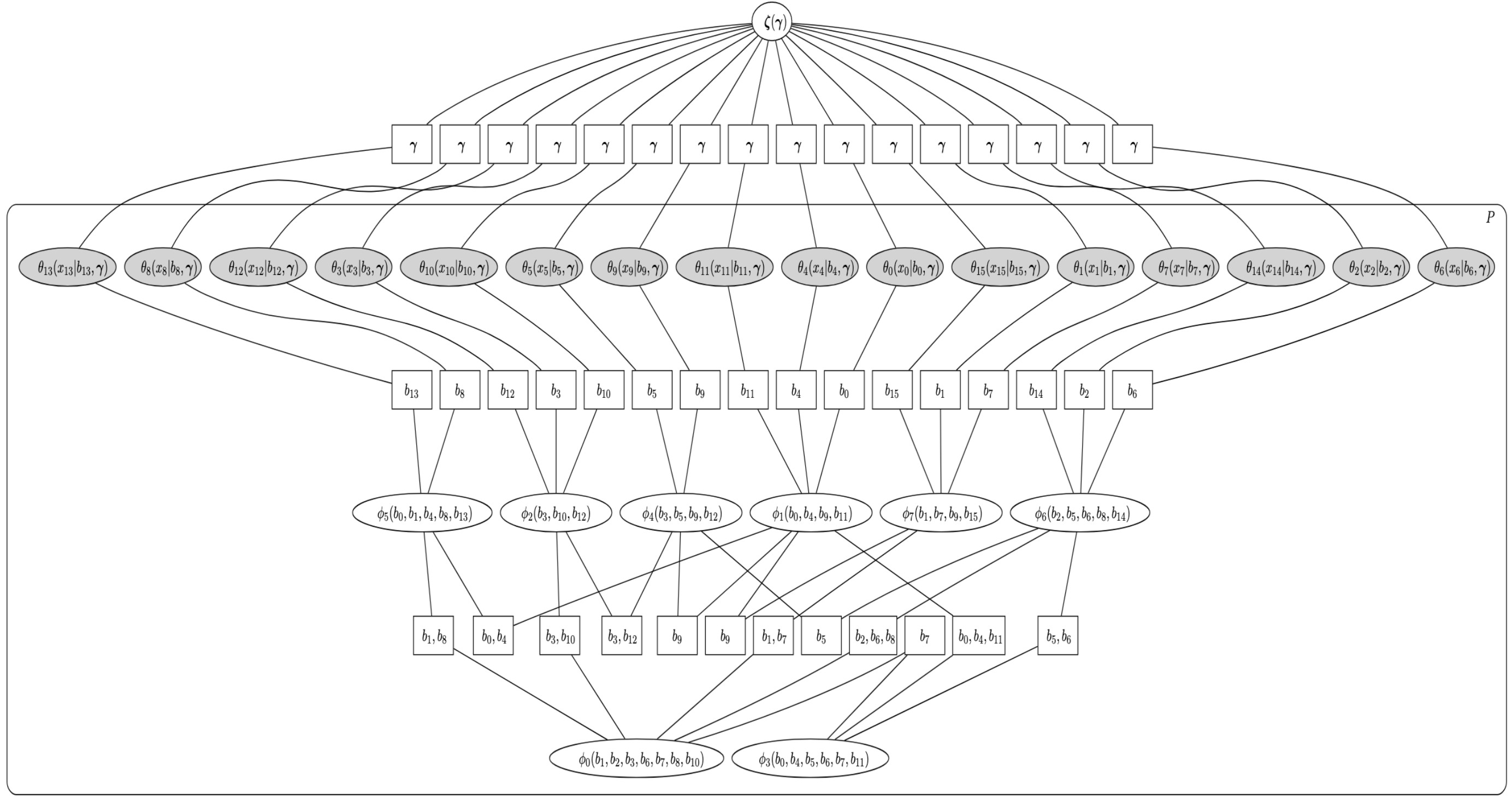}
\caption{A PGM of an irregular (16,8) LDPC code with a global gamma prior and conditional Gaussian clusters linked to the smaller parity-check clusters. Note its loopy structure.}
\label{fig:ldpc16}
\end{figure}

For LDPC codes, each cluster contains a parity-check factor equivalent to a parity-check constraint in the original parity-check matrix. To illustrate this, we use an irregular (16,8) LDPC code with $\mathbf{H}$ matrix given in Equation~\ref{eq:hmatrix}. Note that this LDPC code is for illustrative purposes only, we use larger standardised LDPC codes for our simulations. We denote the message bit sequence as $b_0,...,b_7$, the parity-check bits as $b_8,...,b_{15}$, and the parity-check factors as $\phi_{0},...,\phi_{7}$. The cluster graph of the (16,8) LDPC code is shown in Figure~\ref{fig:ldpc16} in plate notation.

\begin{equation}
\label{eq:hmatrix}
\mathbf{H} =
\begin{bmatrix*}[r]
0 & 1 & 1 & 1 & 0 & 0 & 1 & 1 & 1 & 0 & 1 & 0 & 0 & 0 & 0 & 0 \\
1 & 0 & 0 & 0 & 1 & 0 & 0 & 0 & 0 & 1 & 0 & 1 & 0 & 0 & 0 & 0 \\
0 & 0 & 0 & 1 & 0 & 0 & 0 & 0 & 0 & 0 & 1 & 0 & 1 & 0 & 0 & 0 \\
1 & 0 & 0 & 0 & 1 & 1 & 1 & 1 & 0 & 0 & 0 & 1 & 0 & 0 & 0 & 0 \\
0 & 0 & 0 & 1 & 0 & 1 & 0 & 0 & 0 & 1 & 0 & 0 & 1 & 0 & 0 & 0 \\
1 & 1 & 0 & 0 & 1 & 0 & 0 & 0 & 1 & 0 & 0 & 0 & 0 & 1 & 0 & 0 \\
0 & 0 & 1 & 0 & 0 & 1 & 1 & 0 & 1 & 0 & 0 & 0 & 0 & 0 & 1 & 0 \\
0 & 1 & 0 & 0 & 0 & 0 & 0 & 1 & 0 & 1 & 0 & 0 & 0 & 0 & 0 & 1
\end{bmatrix*}
\end{equation}
\smallskip

The gamma prior belief is captured by the global cluster $\zeta(\gamma)$ (outside the plate) that is fully connected to all the observed conditional Gaussian clusters $\theta_{0}(x_{0}\mid b_{0},\gamma),...,\theta_{15}(x_{15}\mid b_{15},\gamma)$ via the $\gamma$ sepsets. The conditional Gaussian clusters connects to the first layer of parity-check clusters of the LDPC code via sepsets $b_0,...,b_{15}$. The structure inside the plate repeats as LDPC packets $p \in P$ arrive at the receiver.

Our study focuses on irregular LDPC codes as adopted by the 5G new radio (NR) standard by the 3rd Generation Partnership Project (3GPP). Irregular LDPC codes are also known to outperform regular LDPC codes~\cite{johnson2006introducing,mackay2003information}. With regular LDPC codes, all parity-check factors have equal cardinality, which is not the case for irregular LDPC codes. Ideally, the larger parity-check factors in irregular LDPC codes should have minimal connectivity, which helps reduce expensive computation during inference. This is not possible to do in a factor graph representation, as the number of edge connections is equal to the cardinality of each parity-check factor. Instead, we bundle the larger parity-check clusters away from the observed conditional Gaussian clusters using a message passing schedule. The message passing schedule is described in Section~\ref{sec:message-passing-schedule}. Note that no connections are present between observed conditional Gaussian clusters and the large parity-check clusters $\phi_0$ and $\phi_3$ as shown in Figure~\ref{fig:ldpc16}. The observed conditional Gaussian clusters are linked to the smaller parity-check clusters $\phi_1$, $\phi_2$, $\phi_4$, $\phi_5$, $\phi_6$, and $\phi_7$, leading to considerable computational savings.

\subsection{Message passing approach}
This section describes \emph{variational message passing} (VMP) between the gamma and conditional Gaussian nodes, \emph{loopy belief update} (LBU) between the parity-check nodes and a hybrid message passing approach between the conditional Gaussian nodes and parity-check nodes. All the required update equations and rules are discussed. We also discuss the message passing schedule mentioned previously, which alleviates expensive parity-check computation.

\subsubsection{Variational message passing} \label{sec:vmp}
This section concerns the message passing between the various conditional Gaussian clusters and the gamma cluster. VMP makes a distinction between \emph{parent-to-child} messages and \emph{child-to-parent} messages~\cite{winn2005variational}. We deal with the parent-to-child messages that update the conditional Gaussian clusters first.

The gamma distribution and its expected moments are required for the parent-to-child messages from the $\gamma$ cluster to the conditional Gaussian clusters $\theta_n$. We use the following parameterisation\footnote{An alternative parameterisation for the gamma distribution is: $\alpha$ (shape), and $\beta$ (rate), which translates to $\nu = 2\alpha$, and $w = \frac{1}{2\beta}$.} of the gamma distribution:
\begin{equation}
\text{Gam}(\gamma \mid \omega,\nu) = \frac{(\frac{1}{2\omega})^\frac{\nu}{2}}{\Gamma(\frac{\nu}{2})} \gamma^{\frac{\nu}{2} - 1}e^{-\frac{\gamma}{2\omega}} \textrm{,} \\
\label{eq:gamma}
\end{equation}
where $\nu$ is the degrees of freedom (equivalent to the number of observations that convey a ``confidence'' in the mean value), $\omega$ is a scaled precision, and $\Gamma$ is the gamma function.

% \jaco{TODO: maybe mention the characteristics of the Gamma distribution here - the relationship between its mean and its variance described by its parameters. Or mention it in the results?}

In exponential family form transformed to the log-domain this is given as:
\begin{align}
\log(\text{Gam}(\gamma \mid \omega,\nu)) &=
\begin{bmatrix}
-\frac{1}{2\omega} \\
\frac{\nu}{2} - 1
\end{bmatrix}^{\textrm{T}}
\begin{bmatrix}
\gamma \\
\log(\gamma)
\end{bmatrix} -(\log(\Gamma(\frac{\nu}{2}))+ \frac{\nu}{2}\log(2\omega)) \textrm{.}
\label{eq:gamma-exp}
\end{align}
In this form, the left column vector is the natural parameters of the distribution and the right column vector is the sufficient statistics of the distribution. The expected moment vector of the sufficient statistics is~\cite[Appendix B]{bishop2006pattern}:
\begin{align}
\left< \gamma\right>, \left< \log(\gamma) \right> &=
\begin{bmatrix}
\nu \omega \\
\psi(\frac{\nu}{2}) - \log(\frac{1}{2 \omega})
\end{bmatrix}^{\textrm{T}}\textrm{,}
\label{eq:expectation-gamma}
\end{align}
where $\psi$ is the \emph{digamma-function}.

The other parent-to-child messages required to update the conditional Gaussian clusters are from the parity-check clusters. To understand these, we first need the categorical distribution in its exponential family form transformed to the log-domain given by~\cite[Section 2.3.1]{Taylor2023}:
\begin{equation}
\log(p(b \mid \pi_{0},\pi_{1})) =
\begin{bmatrix}
\log(\pi_0) \\
\log(\pi_1)
\end{bmatrix}^{\textrm{T}}
\begin{bmatrix}
\llbracket b=0 \rrbracket \\
\llbracket b=1 \rrbracket
\end{bmatrix} - \log(\sum_{i=0,1} \pi_{i}) \textrm{,}
\label{eq:categorical-exp}
\end{equation}
where $\pi_{i}$ are the unnormalised bit probabilities. The expected
moment vector of the sufficient statistics is:
\begin{align}
\left< \llbracket b=0 \rrbracket \right>, \left< \llbracket b=1 \rrbracket \right>
&=
\begin{bmatrix}
\frac{\pi_0}{\sum_{i=0,1}\pi_{i}} \\
\frac{\pi_1}{\sum_{i=0,1}\pi_{i}}
\end{bmatrix}^{\textrm{T}}\textrm{.}
\end{align}

We choose the gamma distribution prior parameters by considering its mean value, given by $\nu\times\omega$, and we think of $\nu$ as the number of observations that its mean value is based on. Since the gamma distribution is conjugate to the Gaussian distribution, the conditional Gaussian likelihood from Equation~\ref{eq:conditional-gaussian} can be rewritten into a similar form. The Gaussian precision is common to both cases of $b$ and is therefore updated as a single entity (unlike the standard Gaussian mixture model). The exponential family form of the conditional Gaussian distribution transformed to the log-domain is given by:
\begin{align}
\log(p(x_n \mid b_n,\gamma)) &=
\begin{bmatrix}
\sum_{i} \llbracket b_n=i \rrbracket(\gamma \mu_{i}) \\
\sum_{i} \llbracket b_n=i \rrbracket (\frac{-\gamma}{2})
\end{bmatrix}^{\textrm{T}}
\begin{bmatrix}
x_{n} \\
x^{2}_{n}
\end{bmatrix} \nonumber \\
& + \frac{1}{2}\sum_{i} \llbracket b_n=i \rrbracket (\log(\gamma) - \gamma \mu^{2}_{i} - \log(2\pi)) \text{.}
\label{eq:mixture-x}
\end{align}
The $x_n$ terms are the observed signal values and the Gaussian means $\mu_i$ are known and can be replaced by their fixed values $\mu_{0}=-1$ and $\mu_{1}=1$.

During the parent-to-child update from the gamma cluster, the $\gamma$ and $\log(\gamma)$ terms in Equation~\ref{eq:mixture-x} will be replaced with their expected values given by Equation~\ref{eq:expectation-gamma}. This is shown by:
\begin{align}
\mu^{'}_{\zeta,\theta_n} &= [\left< \gamma\right>,\left< \log(\gamma) \right>]^{\text{T}} \text{,} \label{eq:p2c-gamma-message} \\
\Psi^{'}_{\theta_n} &=
\begin{bmatrix}
\sum_{i} \llbracket b_n=i \rrbracket(\left<\gamma\right> \mu_{i}) \\
\sum_{i} \llbracket b_n=i \rrbracket (\frac{-\left<\gamma\right>}{2})
\end{bmatrix}^{\textrm{T}}
\begin{bmatrix}
x_{n} \\
x^{2}_{n}
\end{bmatrix} \nonumber \\
& + \frac{1}{2}\sum_{i} \llbracket b_n=i \rrbracket (\left<\log(\gamma)\right> - \left<\gamma\right> \mu^{2}_{i} - \log(2\pi)) \text{.} \label{eq:p2c-gamma-x-update}
\end{align}

Similarly, during the parent-to-child update from a parity-check cluster the Iverson function $\llbracket b_n=i \rrbracket$ in Equation~\ref{eq:mixture-x} will be replaced with its expected value.

The required expected value from a parity-check cluster reduces to estimating the bit probabilities $P(b_n=i)$. We discuss the replacement of the Iverson function with expected values from parity-check clusters in Section~\ref{sec:hybrid-message-passing} which deals with the hybrid message passing.

The expected values that replace the $\gamma$ and $\llbracket b_n=i \rrbracket$ terms in Equation~\ref{eq:mixture-x} are based on the latest beliefs from the gamma cluster distribution and the parity-check cluster distribution. After the expected values are installed, Equation~\ref{eq:mixture-x} forms a mixture of two gamma distributions since $x_n$ is observed and $\mu_i$ are known value. Equation~\ref{eq:mixture-x} is kept in a conditional form rather than a higher dimensional joint form.

Messages from conditional Gaussian clusters to parity-check clusters are child-to-parent messages. We also postpone discussing these message updates to Section~\ref{sec:hybrid-message-passing} (that deals with hybrid message passing).

Messages from conditional Gaussian clusters to the gamma cluster are child-to-parent messages that require Equation~\ref{eq:mixture-x} to be marginalised towards obtaining gamma distribution parameters. The terms of Equation~\ref{eq:mixture-x} are re-arranged to obtain the appropriate parameters given by:
\begin{align}
\log(p(x_n \mid b_n,\gamma)) &=
\begin{bmatrix}
\sum_{i} \llbracket b_n=i \rrbracket(-\frac{1}{2}(x_{n} - \mu_i)^2) \\
\sum_{i} \llbracket b_n=i \rrbracket (\frac{1}{2})
\end{bmatrix}^{\textrm{T}}
\begin{bmatrix}
\gamma \\
\log(\gamma)
\end{bmatrix} \nonumber \\
& - \sum_{i} \llbracket b_n=i \rrbracket (\log(2\pi))
\label{eq:mixture-gamma} \textrm{.}
\end{align}

We note in Equation~\ref{eq:mixture-gamma} that for child-to-parent messages, the separation of the sufficient statistic vector during marginalisation removes the previous parent-to-child message from the updated natural parameter vector. With $\llbracket b_n=i \rrbracket$ replaced by its expected value, $x_n$ replaced with the observed signal value, and $\mu_{0}=-1$ and $\mu_{1}=1$, the left column gives the incremental update with which the prior gamma parameters need to increase to form the posterior. This is shown by:
\begin{align}
\mu^{'}_{\theta_n,\zeta} &=
\begin{bmatrix}
\sum_{i=0,1} \left<\llbracket b_n=i \rrbracket\right> (-\frac{1}{2}(x^{2}_{n} - 2 x_n \mu_i + \mu^{2}_i) \label{eq:c2p-gamma-mesage} \\
\sum_{i=0,1} \frac{1}{2} \left<\llbracket b_n=i \rrbracket\right>
\end{bmatrix}^{\textrm{T}} \textrm{,} \\
\Psi_{{\zeta}_{\text{prior}}} &=
\begin{bmatrix}
-\frac{1}{2\omega} \\
\frac{\nu}{2} - 1
\end{bmatrix}^{\textrm{T}} \textrm{,} \\
\Psi^{'}_{\zeta} &= \Psi_{{\zeta}_{\text{prior}}} + \sum_{n} \mu^{'}_{\theta_n,\zeta}\textrm{.} \label{eq:c2p-gamma-update}
\end{align}

The expected values of the updated gamma cluster can now be recalculated using Equation~\ref{eq:expectation-gamma} to be used for updating the conditional Gaussian clusters in the next round.

\begin{figure}[!h]
     \centering
     \includegraphics[width=0.55\textwidth]{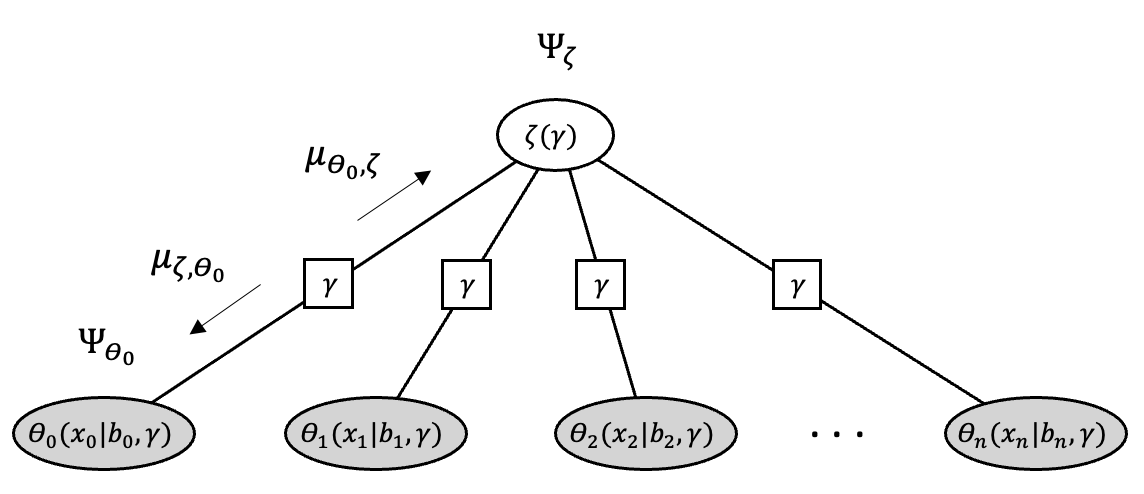}
     \caption{A subsection of Figure~\ref{fig:ldpc16} to illustrate VMP message passing between the gamma cluster and conditional Gaussian clusters.}
     \label{fig:vmp-example}
\end{figure}

With this background, we demonstrate how VMP is applied (using a sub-graph extracted
from Figure~\ref{fig:ldpc16}) in Figure~\ref{fig:vmp-example}.

Cluster $\theta_0(x_0\mid b_0,\gamma)$ is updated with information from cluster $\zeta(\gamma)$ using a VMP parent-to-child message $\mu_{\zeta,\theta_0}$. This message is the expected values of $\log(\text{Gam}(\gamma \mid \omega,\nu))$ given by Equation~\ref{eq:expectation-gamma}. The expected values are installed at the corresponding $\gamma$ terms in Equation~\ref{eq:mixture-x}. The message from cluster $\theta_0(x_0\mid b_0,\gamma)$ to cluster $\zeta(\gamma)$ is a VMP child-to-parent message $\mu_{\theta_0,\zeta}$. This requires cluster $\theta_0(x_0\mid b_0,\gamma)$ to be marginalised in obtaining a distribution over sepset $\gamma$ with its parameters given by the natural parameter vector in Equation~\ref{eq:mixture-gamma}. The message is absorbed into cluster $\zeta(\gamma)$ by adding its natural parameters to those of the prior (as given in Equation~\ref{eq:gamma-exp}).

\subsubsection{LBU message passing}
\label{sec:lbu-message-passing}
This section concerns the message passing between the various
parity-check clusters -- here we make use of the
Lauritzen-Spiegelhalter message passing algorithm
~\cite{Lauritzen1988}, also known as the \emph{loopy belief update}
(LBU) algorithm. However, the fundamental concepts are easier to
define via the well known Shafer-Shenoy algorithm ~\cite{Shafer1990},
also known as the sum-product or \emph{loopy belief propagation} (LBP)
algorithm. Hybrid message passing, linking the VMP and the LBU
sections of the graph, also draws on understanding the relationship
between LBP and LBU. We therefore provide a brief summary of both LBP
and LBU here -- a fuller version is available in the definitive handbook
by Koller \& Friedman \cite[Sections 10.2, 10.3 and 11.3]{Koller2009}.

\begin{figure}[!h]
     \centering
     \includegraphics[width=0.6\textwidth]{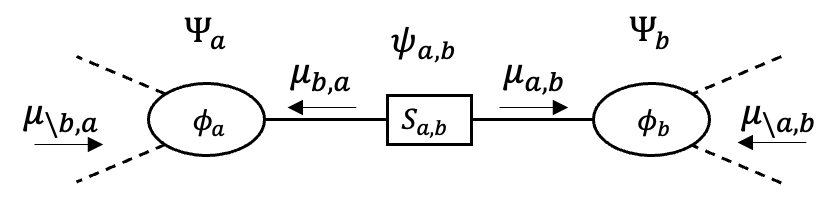}
     \caption{Example cluster graph to illustrate LBP and LBU message updating rules.}
     \label{fig:lbu-messages}
\end{figure}

We use Figure~\ref{fig:lbu-messages} to illustrate the various
relevant concepts. As discussed in Section~\ref{sec:representation},
cluster graphs contain cluster nodes and sepsets, both of which are
used during message passing to update nodes. The cluster internal
factor functions are given by $\phi_a$ and $\phi_b$ where $a$ and $b$
identifies the particular cluster (in our application these functions
are conditional distributions enforcing even parity over all the
variables involved in each parity-check cluster). We will at times
somewhat abuse the notation by also using these factor functions
directly to identify their particular clusters.

The sepset connecting the two clusters $a$ and $b$ is denoted as
$\mathcal{S}_{a,b}$ and comprises the collection of random variables
about which the two clusters will exchange information. The notation
$\backslash a$ is used to indicate the set of all cluster nodes
excluding cluster node $a$.

For LBP message passing, the message passed from cluster $a$ to
cluster $b$ is denoted as $\mu_{a,b}$. The product of all other
messages incoming to cluster node $a$ -- excluding message $\mu_{b,a}$
-- is denoted by $\mu_{\backslash b,a}$. If anything changes in the
messages $\mu_{\backslash b,a}$, we can propagate this towards cluster
$b$ via the update equation:
\begin{equation}
  \mu^{'}_{a,b} = \sum_{\backslash\mathcal{S}_{a,b}} \mu_{\backslash b,a} \phi_{a} \text{,}
  \label{eq:mu_lbp}
\end{equation}
where the marginalisation sum over the set
$\backslash\mathcal{S}_{a,b}$ removes all variables not present in the
sepset $\mathcal{S}_{a,b}$. Note that this implies that the larger a
sepset, the cheaper the required marginalisation will be -- another
benefit of cluster graphs over factor graphs which always have only
single variables in their sepsets.

Also note that, especially with clusters sharing links with many
others, the LBP formulation can be quite expensive due to the
redundancy implied by the various message combinations present in the
$\mu_{\backslash b,a}$ term.

In a loopy system (such as we will typically have with LDPC codes)
these $\mu$ messages will have to be passed iteratively according to
some schedule until they have all converged. At that point, we can then
get an estimate for the marginal distribution of all the random
variables present in a particular cluster -- this is known as the
cluster belief and is given by:
\begin{equation}
  \Psi_{a} =  \phi_{a} \mu_{b,a} \mu_{\backslash b,a} \text{,}
  \label{eq:Psi_lbp}
\end{equation}
i.e. it is the product of the cluster internal factor with \emph{all}
incoming messages.  Similarly, we can calculate the belief over the
sepset variables $\mathcal{S}_{a,b}$ as the product of the two
opposing messages passing through that sepset:
\begin{equation}
    \psi_{a,b} =  \mu_{a,b} \mu_{b,a} \text{.}
\end{equation}
Note that the sepset beliefs, being the product of the two $\mu_{a,b}$
and $\mu_{b,a}$ messages, are intrinsically directionless.

In contrast to LBP, LBU message passing is expressed fully in terms of
only \emph{cluster beliefs} and \emph{sepset beliefs}. For this we use
two alternative (although equivalent) expressions for these quantities
(see \cite[Section 10.3]{Koller2009} for the derivations). The sepset
belief update is given by:
\begin{align}
  \psi^{'}_{a,b} &= \sum_{\backslash\mathcal{S}_{a,b}} \Psi^{'}_a \text{.} \label{eq:psi_lbu}
\end{align}
Note that this is computationally more efficient since it avoids the
repeated re-combination of the various messages present in the
$\mu_{\backslash b,a}$ term in Equation~\ref{eq:mu_lbp} when
considering different target clusters $b$. Using this we can now
incrementally update the cluster belief using:
\begin{align}
  \Psi^{'}_{b} &= \Psi_b\frac{\psi^{'}_{a,b}}{\psi_{a,b}} \text{.} \label{eq:Psi_lbu}
\end{align}

\begin{figure}[!h]
     \centering
     \includegraphics[width=0.93\textwidth]{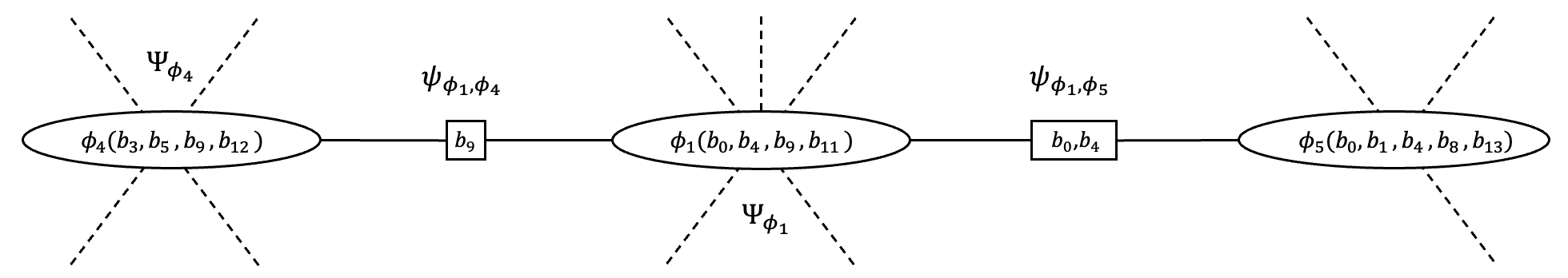}
     \caption{A subsection of Figure~\ref{fig:ldpc16} to illustrate
       LBU message passing between parity-check clusters. The dotted
       lines denote links to other adjacent clusters.}
     \label{fig:lbu}
\end{figure}

With this background, we can now turn to Figure~\ref{fig:lbu} (a
sub-graph extracted from Figure~\ref{fig:ldpc16}) to illustrate how
LBU applies to the parity-check clusters.
Cluster node $\phi_4$ is updated using LBU message passing. Using
Equation~\ref{eq:psi_lbu}, the updated sepset belief is calculated by
marginalising the cluster belief $\Psi_{\phi_1}$ to obtain the
required sepset $b_9$. The previous sepset belief is cancelled
(divided) as per the LBU update rules given by
Equation~\ref{eq:Psi_lbu}. This gives the update equations as:
\begin{align}
\psi^{'}_{\phi_1,\phi_4} &= \sum_{b_0,b_4,b_{11}} \Psi_{\phi_1} \text{,} \\
 \Psi^{'}_{\phi_4} &= \Psi_{\phi_4} \frac{\psi^{'}_{\phi_1,\phi_4}}{\psi_{\phi_1,\phi_4}} \text{.}
\end{align}

Unlike VMP, the form of the update rules remains the same regardless
of the message direction. We therefore simply iterate these messages
until convergence using the message passing schedule of
Section~\ref{sec:message-passing-schedule}.

\subsubsection{Hybrid message passing}
\label{sec:hybrid-message-passing}
\begin{figure}[h]
\centering
\includegraphics[width=0.5\textwidth]{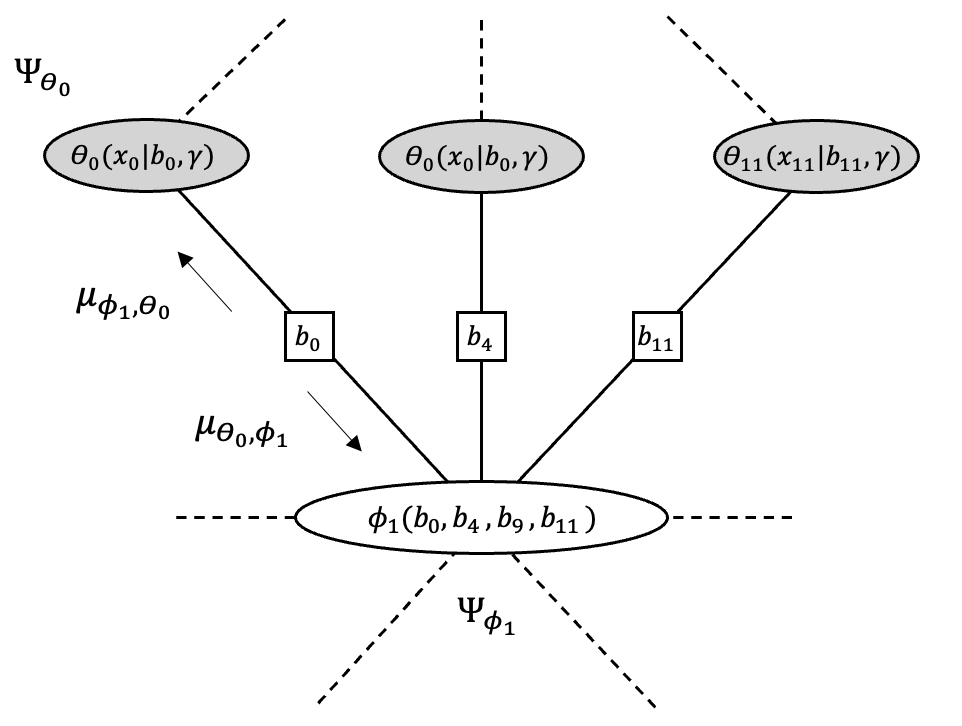}
\caption{A subsection of Figure~\ref{fig:ldpc16} to illustrate hybrid message passing and updates between conditional Gaussian clusters and parity-check clusters.}
\label{fig:lbu-vmp}
\end{figure}

This section describes our hybrid message passing between the
VMP-based conditional Gaussian cluster nodes and the LBU-based
parity-check nodes connected to it. Although our treatment is general
and applies to any $\{b_n\}$, we will specifically focus on the
$\mu_{\theta_0,\phi_1}$ and $\mu_{\phi_1,\theta_0}$ messages running
over the $\{b_0\}$ sepset in Figure~\ref{fig:lbu-vmp} (which in its
turn is a sub-graph extracted from Figure~\ref{fig:ldpc16}).

We first consider the child-to-parent message
$\mu^{'}_{\theta_0,\phi_1}$ running from the $\theta_0(x_0\mid
b_0,\gamma)$ cluster to the $\phi_1(b_0,b_4,b_9,b_{11})$
cluster. According to VMP message passing we determine its natural
parameters by re-arranging Equation~\ref{eq:mixture-x} so that the
$b_0$ related terms move to the sufficient statistics column:

\begin{align}
  \log(p(x_0 \mid b_0,\gamma)) &=
  \begin{bmatrix}
    \gamma \mu_{0}x_0 - \frac{\gamma x_0^{2}}{2} + \frac{(\log(\gamma) - \gamma \mu_{0}^{2} - \log(2\pi))}{2} \\
    \gamma \mu_{1}x_0 - \frac{\gamma x_0^{2}}{2} + \frac{(\log(\gamma) - \gamma \mu_{1}^{2} - \log(2\pi))}{2}
  \end{bmatrix}^{\textrm{T}}
  \begin{bmatrix}
    \llbracket b_0=0 \rrbracket \\
    \llbracket b_0=1 \rrbracket
  \end{bmatrix} \textrm{.}
  \label{eq:categorical}
\end{align}
From our model definition, the values of the means are known as
$\mu_0=-1$ and $\mu_1=1$. The expected values of the $\gamma$ terms
are known via Equation~\ref{eq:expectation-gamma}. The natural
parameters column (the left-hand one) effectively measure the heights
of the two Gaussians corresponding to $b_0$ being either $0 \textrm{
  or } 1$. Similar to what we saw before with
Equation~\ref{eq:mixture-gamma}, we note that for child-to-parent
messages, the separation of the sufficient statistic vector during
marginalisation effectively \emph{removes} the previous
parent-to-child message from the updated natural parameter vector.

The question now is how this is to be made compatible with the
corresponding $\phi_1$ cluster which, of course, will also be updated
via sepset beliefs from its other side (connecting to other
parity-check clusters using Equation~\ref{eq:Psi_lbu}). To reconcile
these worlds, we start by comparing Equations~\ref{eq:mu_lbp} and
\ref{eq:Psi_lbp} and noticing that we can also determine the
(original) update message using:
\begin{equation}
  \mu^{'}_{a,b} = \sum_{\backslash\mathcal{S}_{a,b}} \Psi_a / \mu_{b,a} \text{.}
  \label{eq:mu_lbu2}
\end{equation}
Due to that (explicit) division, we refer to this as the
\emph{post-cancellation} message, i.e. the message we derive by
marginalising the cluster belief\footnote{From
Equation~\ref{eq:psi_lbu} this is the sepset belief -- we also term it
the \emph{pre-cancellation} message.} and then removing the opposite
direction message via division. But this strongly reminds us of the
child-to-parent message of Equation~\ref{eq:categorical} which also
(implicitly) accomplishes removing the influence of its parent. We
therefore set our updated $\mu^{'}_{\theta_0,\phi_1}$ to be a
a categorical message with natural parameters given by the corresponding
lefthand column of Equation~\ref{eq:categorical}.

Next, we need to determine how to incrementally update the cluster
belief $\Psi_{\phi_1}$ with such a post-cancellation message. We do
this via Equation~\ref{eq:Psi_lbp} (but now with $\Psi_b$ in mind) to
get the cluster belief update equation:
\begin{align}
  \Psi^{'}_{b}
  &=  \phi_{b} \mu^{'}_{a,b} \mu_{\backslash a,b} \text{,} \nonumber \\
  &=   \Psi_{b} \frac{\mu^{'}_{a,b}}{\mu_{a,b}} \text{.}
  \label{eq:Psi_lbu2}
\end{align}
In short, we can directly update an LBU cluster belief by dividing the
old post-cancellation message out and multiplying the new one in. This
gives us:
\begin{align}
  \Psi^{'}_{\phi_{1}} &= \Psi_{\phi_{1}}\frac{\mu^{'}_{\theta_0,\phi_1}}{\mu_{\theta_0,\phi_1}}\text{.}
\end{align}

The child-to-parent update rules for the general case that connects a
conditional Gaussian cluster $\theta_n$ to a parity-check cluster
$\phi_j$ are:
\begin{align}
  \mu_{\theta_n,\phi_j} &= \begin{bmatrix}
    \gamma \mu_{0}x_n - \frac{\gamma x_n^{2}}{2} + \frac{(\log(\gamma) - \gamma \mu_{0}^{2} - \log(2\pi))}{2} \\
    \gamma \mu_{1}x_n - \frac{\gamma x_n^{2}}{2} + \frac{(\log(\gamma) - \gamma \mu_{1}^{2} - \log(2\pi))}{2}
  \end{bmatrix}^{\textrm{T}} \text{,} \label{eq:c2p-b-message} \\
\Psi^{'}_{\phi_{j}} &= \Psi_{\phi_{j}}\frac{\mu^{'}_{\theta_n,\phi_j}}{\mu_{\theta_n,\phi_j}}\text{.} \label{eq:c2p-b-update}
\end{align}

Lastly, we consider the parent-to-child message
$\mu^{'}_{\phi_1,\theta_0}$ running from the
$\phi_1(b_0,b_4,b_9,b_{11})$ cluster to the $\theta_0(x_0\mid
b_0,\gamma)$ cluster. From the VMP viewpoint, our message
$\mu^{'}_{\phi_1,\theta_0}$ simply is the expected values:
\begin{align}
  \mu^{'}_{\phi_1,\theta_0}
  &= \left<\llbracket b_0=i\rrbracket\right>_{\Psi_{\phi_1}(b_0,b_4,b_9,b_{11})} && \textrm{with } i \in \{0,1\}\nonumber.
\end{align}
The first step is to find the distribution over $b_0$ by marginalising:

\begin{align}
  \psi_{\theta_0,\phi_1}(b_0) &= \sum_{b_4,b_9,b_{11}} \Psi_{\phi_1}(b_0,b_4,b_9,b_{11}).
\label{eq:expectation-iverson}
\end{align}
The required expectations are the probabilities for $b_0$ as found in
this sepset belief - we simply install them as the required parameters
in the $\theta_0$ cluster.

In general, where we have a message describing $b_n$ running from the
parent cluster $\phi_j$ to a child cluster $\theta_n$, we first find
the sepset belief:
\begin{align}
  \psi^{'}_{\theta_n,\phi_j}(b_n) &= \sum_{\backslash b_n} \Psi_{\phi_j} \text{,} \label{eq:p2c-b-message}
\end{align}
and then replace the two probabilities for $b_n$ found in this sepset
belief as the required expected sufficient statistics $\left<\llbracket
b_n=i\rrbracket\right>_{\Psi_{\phi_j}}$ in the $\theta_n$ cluster:
\begin{align}
\Psi^{'}_{\theta_n} &=
\begin{bmatrix}
\sum_{i} \left<\llbracket b_n=i \rrbracket\right>_{\Psi_{\phi_{j}}} (\gamma \mu_{i}) \\
\sum_{i} \left<\llbracket b_n=i \rrbracket\right>_{\Psi_{\phi_{j}}} (\frac{-\gamma}{2})
\end{bmatrix}^{\textrm{T}}
\begin{bmatrix}
x_{n} \\
x^{2}_{n}
\end{bmatrix} \nonumber \\
& + \frac{1}{2}\sum_{i} \left<\llbracket b_n=i \rrbracket\right>_{\Psi_{\phi_{j}}} (\log(\gamma) - \gamma \mu^{2}_{i} - \log(2\pi)) \text{.} \label{eq:p2c-b-update}
\end{align}

\subsubsection{Message passing schedule}
\label{sec:message-passing-schedule}
A message passing schedule is important for loopy graphs as the message order can influence convergence speed, accuracy, and the computational cost of inference. Although not empirically verified here, these feedback loops (or cycles) may reinforce inaccurate cluster beliefs causing self-fulfilling belief updates, which affect the LDPC decoder's performance. This problem is more prominent in LDPC codes with small feedback loops as described in~\cite{johnson2006introducing}. Taking this into consideration, our message passing schedule (1) uses a structured schedule with a fixed computational cost, (2) aims to minimise the effect of loops, and (3) aims to minimise the computational cost of inference.

The message passing schedule is determined by first identifying the larger parity-check clusters in the graph. We select the larger clusters $\phi_0$ (with cardinality 7) and $\phi_3$ (with cardinality 6). The message schedule starts with the selected clusters as initial sources $\mathcal{S}$ and proceeds by visiting all its neighbouring clusters $\mathcal{N}$, which become the immediate next layer of clusters. A set of available clusters $\mathcal{A}$ is kept to make sure that clusters from previous layers are not revisited, which helps minimise the effect of loops. We repeat this procedure to add subsequent layers of clusters until all clusters are included. The source-destination pairs are stored in a message schedule $\mathcal{M}$. This procedure isolates the initially selected large parity-check clusters from the rest of the clusters as shown in Figure~\ref{fig:ldpc16}. The idea is to keep the expensive clusters at the final layer so that the smaller (less expensive) parity clusters, in preceding layers, can resolve most of the uncertainty about the even parity states. When the larger parity clusters get updated, some of the even parity states in their discrete tables may have zero probability, which are removed due to our software implementation. This further reduces a large parity cluster's computational footprint. Our layered message passing schedule is detailed in Algorithm 1.
\begin{algorithm}
\caption{Layered message passing schedule}\label{alg:alg1}
\begin{algorithmic}[1]
\State $\mathcal{S} \gets$ set initialised to large cluster IDs
\State $\mathcal{A} \gets$ set initialised to all cluster IDs
\State $\mathcal{M} \gets$ empty vector of pairs
\While{$\mathcal{A}$ is not empty}
\State $nextLayer \gets$ empty set of integers
\For{$s$ in $\mathcal{S}$}
\State $\mathcal{A}$.erase($s$)
\State $\mathcal{N} \gets$ all neighbours of $s$
\For{$n$ in $\mathcal{N}$}
\If{$n \in \mathcal{A}$}
\State $\mathcal{M}$.push\_back(pair($s$, $n$))
\If{$n \not\in \mathcal{S}$}
\State $nextLayer$.insert($n$)
\EndIf
\EndIf
\EndFor
\EndFor
\State $\mathcal{S} \gets nextLayer$
% \STATE $nextLayer$.clear()
\EndWhile
\State
\Return $\mathcal{M}$
\end{algorithmic}
\label{alg1}
\end{algorithm}

The observed conditional Gaussian clusters are coupled to the parity-check clusters in the layer furthest away from the initial isolated group of large clusters. We refer to this layer as the first parity-check layer. The smaller parity-check clusters in this layer are given priority in terms of their connectivity to the conditional Gaussian clusters, which saves computation. If the first layer of parity-check clusters does not have all the unique bits, the following layers are utilised until all conditional Gaussian clusters are connected.

The message passing order for the entire PGM starts at the gamma cluster and flows down to the conditional Gaussian clusters through all the parity cluster layers until it reaches the bottom layer. We refer to this as the forward sweep. The backward sweep returns in the opposite direction, which concludes one iteration of message passing. Pseudo-code for our message passing is shown in Algorithm~\ref{alg:alg2}.
\begin{algorithm}[H]
\caption{Message passing}\label{alg:alg2}
\begin{algorithmic}[1]
\State $\psi \gets$ initialised to uniform sepset beliefs
\State $\mathcal{M} \gets$ initialised to source-destination pairs from message schedule
\State $\mathcal{S}$ initialised to sepset variable sets between clusters
\State $\Psi_{\zeta_{prior}} \gets$ initialised to gamma prior parameters
\State $maxIter \gets$ initialised to maximum number of sweeps
\While{$iter < maxIter$ and not converged}
\State //   --------------  Forward sweep  --------------
\State //VMP messages from cluster $\zeta$ to clusters $\theta_{n}$
\For{each conditional Gaussian cluster $n$}
\State $\mu^{'}_{\zeta,\theta_n} \gets$ from Equation~\ref{eq:p2c-gamma-message}
\State $\Psi^{'}_{\theta_n} \gets$ from Equation~\ref{eq:p2c-gamma-x-update}
\EndFor
\State //Hybrid messages from $\theta_{n}$ to first layer parity-check clusters $\phi_j$
\For{each conditional Gaussian cluster $n$}
\State $\mu^{'}_{\theta_n,\phi_j} \gets$ from Equation~\ref{eq:c2p-b-message}
\State $\Psi^{'}_{\phi_j} \gets$ from Equation~\ref{eq:c2p-b-update}
\EndFor
\State //LBU messages from first layer to final layer parity-check clusters
\State //$s$ is parity-check source cluster, $d$ is parity-check destination cluster
\For{pair($s,d$) in $\mathcal{M}$}
\State $\psi^{'}_{\phi_{s}, \phi_{d}} \gets$ from Equation~\ref{eq:psi_lbu}
\State $\Psi^{'}_{\phi_{d}} \gets$ from Equation~\ref{eq:Psi_lbu}
\State $\psi_{\phi_{s},\phi_{d}} \gets \psi^{'}_{\phi_{s}, \phi_{d}}$
\EndFor
% \State // Continues below
% \algstore{myalg}
% \end{algorithmic}
% \end{algorithm}
% \begin{algorithm}[H]
% \begin{algorithmic}[1]
% \algrestore{myalg}
\State //   --------------  Backward sweep  --------------
\State //LBU messages from final layer towards first layer clusters $\phi_j$
\For{pair($d,s$) in Reverse($\mathcal{M}$)}
\State $\psi^{'}_{\phi_{s}, \phi_{d}} \gets$ from Equation~\ref{eq:psi_lbu}
\State $\Psi^{'}_{\phi_{d}} \gets$ from Equation~\ref{eq:Psi_lbu}
\State $\psi_{\phi_{s},\phi_{d}} \gets \psi^{'}_{\phi_{s}, \phi_{d}}$
\EndFor
\State //Hybrid messages from first layer parity-check clusters $\phi_j$ to $\theta_n$
\For{each conditional Gaussian cluster $n$}
\State $\psi^{'}_{\theta_n,\phi_j}(b_n) \gets$ from Equation~\ref{eq:p2c-b-message}
\State $\Psi^{'}_{\theta_n} \gets$ from Equation~\ref{eq:p2c-b-update}
\EndFor
\State $\Psi_{\zeta} \gets$ set to uniform prior
\State //VMP messages from $\theta_n$ to $\zeta$
\For{each conditional Gaussian cluster $n$}
\State $\mu_{\theta_n,\zeta} \gets$ from Equation~\ref{eq:c2p-gamma-mesage}
\State $\Psi_{\zeta} \mathrel{+}= \mu_{\theta_n,\zeta}$
\EndFor
\State $\Psi^{'}_{\zeta} = \Psi_{\zeta_{prior}} + \Psi_{\zeta}$ (see Equation~\ref{eq:c2p-gamma-update})
\State $iter++$
\If{clusters $\phi_j$ calibrated and codeword valid}
\State converged
\EndIf
\EndWhile
\end{algorithmic}
\label{alg2}
\end{algorithm}

The following settings and software implementations apply to our inference approach:
\begin{itemize}
    \item a cluster is deactivated during message passing when messages entering it have not changed significantly. This is determined by a symmetrical\footnote{The average divergence from both directions is used, taking into consideration the mode-seeking and moment-seeking behaviours of the metric given by $\frac{D_{KL}(P||Q) + D_{KL}(Q||P)}{2}$.} Kullback-Leibler divergence measure between the newest and immediately preceding sepset beliefs.
    \item the stopping criterion for inference is when a valid codeword was detected (also known as a syndrome check) after all parity-check clusters ``agree'' on their shared bit values or when a maximum number of iterations is reached,
    \item all discrete table factors support sparse representations to reduce memory resources,
    \item zero probability states in discrete tables are removed during inference.
\end{itemize}

The next section describes how the LDPC code tracks non-stationary SNRs using a Bayesian sequential learning technique.

\section{Bayesian sequential learning for non-stationary SNR estimation}
\label{sec:bayesian-sequential}
Channel noise in a wireless communication system can change over time, especially when the end user is mobile. This means the statistical properties of the received signal should be treated as \emph{non-stationary}. For the PGM to remain representative of the varying channel conditions, the parameters of the gamma distribution need to adapt accordingly as LDPC packets arrive at the receiver. This can be achieved with Bayesian sequential learning also known as Bayesian sequential filtering. Each time an LDPC packet is decoded, the parameters of the obtained posterior gamma distribution are stored and used as the prior distribution when decoding the next LDPC packet. However, as an increased number of LDPC packets are decoded, our stationarity assumptions cause an ever-increasing certainty around the gamma distribution's mean and the PGM struggles to respond accurately to the changing noise. Tracking the noise becomes restricted due to the strong underlying i.i.d. assumption across all noise estimates (from the observed data), which will follow the average channel noise instead of the evolving channel noise.

To remedy this, we introduce a time constant $T$ that represents a period of time in which we assume the stationarity assumption holds. For example, if we assume no drastic changes in channel noise over a period of $T$ seconds, our stationarity assumption should be valid for $S=\frac{T\times1000}{1}$ number of LDPC packets (assuming a transmission time of 1ms per LDPC packet).

We ``relax'' the i.i.d. assumption by accumulating the posterior gamma distribution's parameters as we normally would, but only up to $S$ number of LDPC packets. After the $S$ number of packets are decoded, the posterior gamma distribution's parameters (the natural parameter vector) are continuously being re-scaled with a scaling factor $\frac{S\times N}{\frac{\nu}{2} - 1}$ as follows:
\begin{equation}
        \Psi^{'}_{\zeta} =
\begin{cases}
    \begin{bmatrix}
    -\frac{1}{2\omega} \\
    \frac{\nu}{2} - 1
    \end{bmatrix}^{\textrm{T}} , & \hspace{8mm} \frac{\nu}{2} - 1 \leq S\times N \\
        \begin{bmatrix}
    -\frac{1}{2\omega} \\
    \frac{\nu}{2} - 1
    \end{bmatrix}^{\textrm{T}} \frac{S\times N}{\frac{\nu}{2} - 1}, & \hspace{8mm} \frac{\nu}{2} - 1 > S\times N \text{.}
\end{cases}
\end{equation}

By scaling the gamma distribution's parameters, the number of observations $\nu$ that make up the posterior distribution reaches a ceiling value, which is not equal to the theoretical maximum of the full i.i.d. assumption. This allows for variance around the gamma distribution's mean so that the estimate does not become too ``confident'' in the data, which makes the gamma distribution responsive to the evolving channel noise. Scaling the natural parameter vector in this way makes contributions from previous estimates progressively less important compared to more recent estimates -- allowing the PGM to ``forget'' historic channel noise. Note that this is not a windowing approach or a Kalman filter type approach where noise is added between estimations. The posterior gamma distribution remains informed by the entire sequence of received data, but the contributions from past data decay exponentially as packets are received and decoded.

\section{Experimental investigation}
\label{sec:experimental-investigation}
This section describes the results obtained from stationary noise and non-stationary noise experiments.

The LDPC code used is constructed from the 5G new radio (NR) standard~\cite{bae2019overview}. We use base graph 2 with size (42, 52) and expansion factor 11. The resultant $\mathbf{H}$ matrix is shortened to a codeword length of $N=220$ bits, of which $K=110$ are message bits (producing a code rate $R=0.5$). A cluster graph is compiled from the parity-check factors using the LTRIP algorithm, and the message schedule is initialised by clusters with cardinality 8 and 10 (the largest factors), which form the bottom layer of the PGM (see Section~\ref{sec:message-passing-schedule}). We use BPSK modulation over an AWGN channel and assume a transmitted bit has unit energy.

\subsection{Purpose of stationary noise experiment}
The stationary noise experiment is presented as a BER vs SNR curve as typically would be the case for determining the behaviour and performance of error correction codes over a range of SNR values. The purpose of the experiment is to compare the BER performance between our proposed PGM and a PGM with perfect knowledge of the noise across a range of SNR values. The selected SNR range consists of 6 equidistant points from 0 to 4.45 dB (inclusive). A random bit message is encoded and Gaussian noise is added to each bit, which repeats for 50k packets per SNR value. The same received bit values are presented to both PGMs to ensure a like-for-like comparison. We set the maximum number of message passing iterations to 20 for both PGMs. Our channel estimation PGM is parameterised with $S = 10$, which assumes stationarity over 10 LDPC packets (equivalent to a period of 10ms).

\subsubsection{Results and interpretation}
\begin{figure}[!h]
     \centering
     \includegraphics[width=\textwidth]{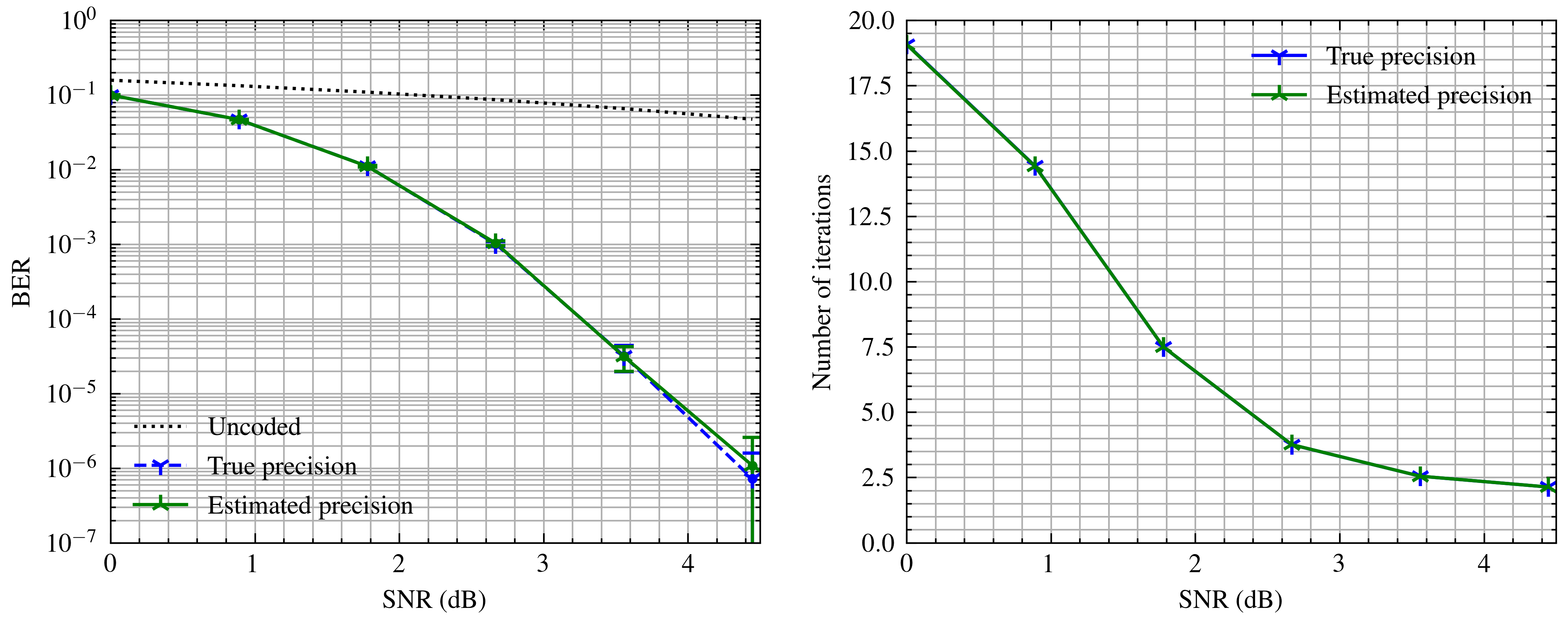}
     \caption{BER performance comparison between our proposed channel noise estimation PGM and a PGM with perfect knowledge of the channel noise. The BER and the number of iterations till convergence between the two systems are nearly identical.}
     \label{fig:ber-test}
\end{figure}

The results of the stationary noise experiment are shown in Figure~\ref{fig:ber-test}. The BER performance (shown on the left) of our proposed PGM closely follows the performance of a PGM with perfect knowledge of the noise precision (the error bars reflect the 95\% confidence intervals). Similarly, the number of iterations required to reach convergence (shown on the right) is almost identical between the two PGMs.

\subsection{Purpose of non-stationary noise experiment}
\label{sec:non-stationary-experiment}
The purpose of the experiment is to test whether our proposed PGM is capable of tracking non-stationary channel noise, and if it benefits from the estimated channel noise information. We compare three scenarios: (1) a PGM that has perfect (instantaneous) knowledge of the channel noise precision, (2) a PGM using a fixed value for the channel noise precision, and (3) our proposed PGM that sequentially updates its channel noise precision estimation. Note that the assumed channel noise in (2) is the last precision mean obtained by running our proposed PGM over the test data once with $S = \infty$. While (2) assumes stationarity, its fixed value of the channel was accumulated from the entire test sequence and is advantageous, since it allows the PGM to ``anticipate'' future values of the channel noise. We set the maximum number of message passing iterations to 20 for all scenarios and parameterise (3) with $S = 10$, which assumes stationarity over 10 LDPC packets (equivalent to a period of 10ms).

\subsubsection{Drive-test data}
\begin{figure}[!h]
     \centering
     \includegraphics[scale=0.40]{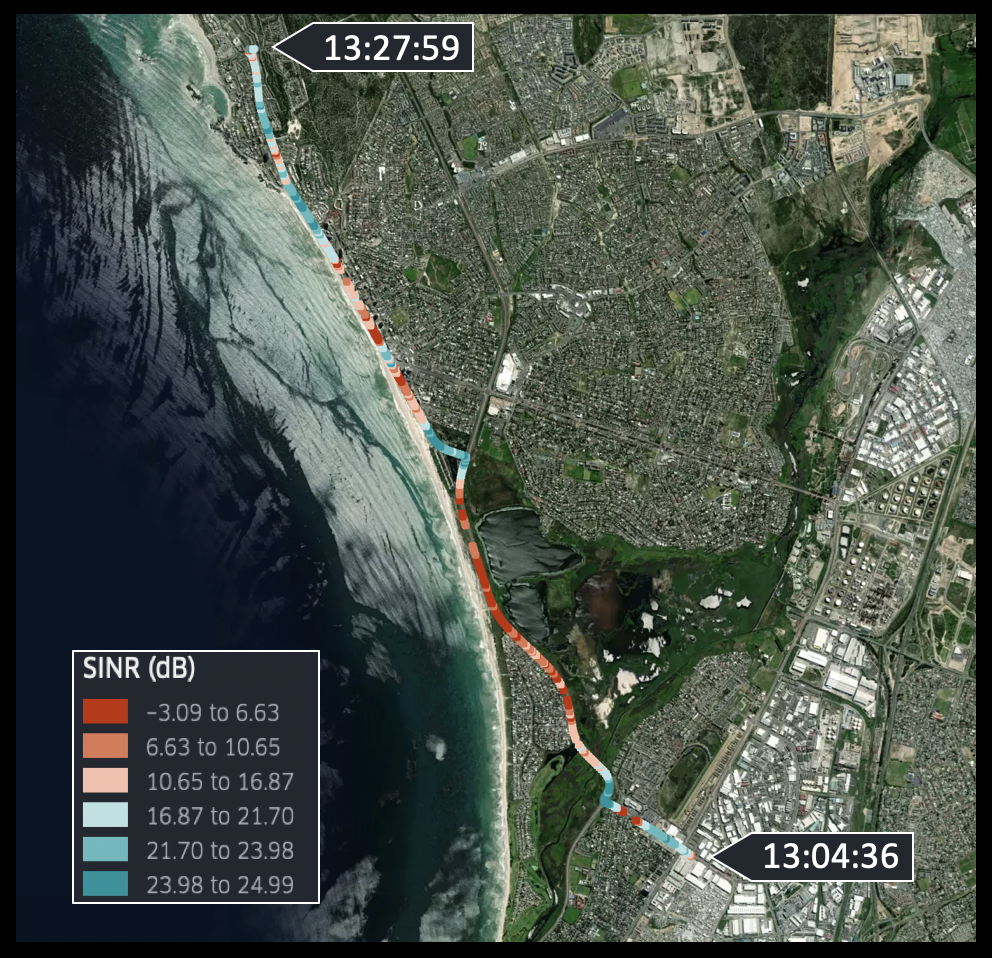}
     \caption{Location of the drive test with the captured SINR measurements. Note the changes in signal-to-noise power as the user equipment moves through certain areas.}
     \label{fig:drive-test}
\end{figure}

We test our model using actual drive test measurements obtained from Vodacom Group Limited, a mobile communications company based in South Africa. Drive tests capture key network performance indicators that allow the mobile network operator to troubleshoot and optimise the network. We use the signal-to-interference-plus-noise ratio (SINR) measurements from a 5G device while driving approximately 10 kilometres along a coastal route. The captured data includes handovers between cells as the user equipment travels in and out of reception range. The start and end times of the test along with the SINR measurements and GPS locations are shown in Figure~\ref{fig:drive-test}.

The drive-test data had missing measurements and irregular time intervals. The missing values were removed and the sequence was re-indexed to create a time series with regular intervals. It was upsampled using linear interpolation to match a basic transmission time interval of 1ms. Note that our final test data may reflect more extreme channel variations due to some discontinuities introduced as a consequence of the described data cleaning.
We generate random bit messages that are encoded using the $\mathbf{H}$ matrix to produce LDPC packets (or codewords). Bit values in an LDPC packet are modulated using BPSK modulation, and random Gaussian noise is added to the resultant signal values using the SINR drive-test data. The SINR data are converted from their logarithmic form to a linear form to obtain precision values. The precision values are used to add zero-mean Gaussian noise to the transmitted signal, which produces the received signal. Note that the same precision value is used to add noise to all signal values from the same LDPC packet.

The dataset used during the current study is not publicly available due to the organisation's data privacy policy, but can be made available from the corresponding author on reasonable request.

\subsubsection{Results and interpretation}
\begin{figure}[!h]
     \centering
     \includegraphics[width=\textwidth]{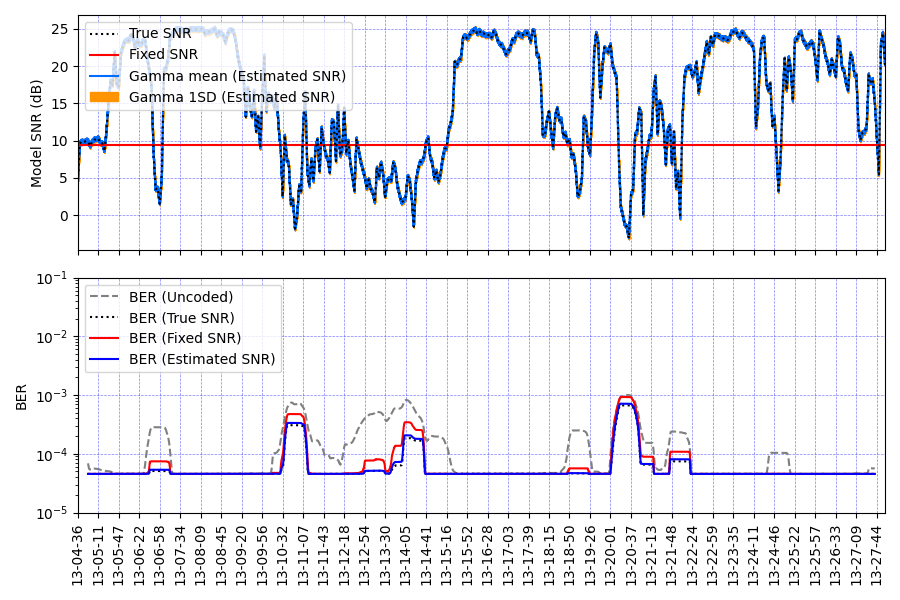}
     \caption{Results of the three scenarios using the 5G device drive-test data. Note that our PGM tracks the changes in precision and maintains a BER similar to that of a PGM with perfect knowledge of the channel. It also outperforms a PGM employing a fixed average precision.}
     \label{fig:results}
\end{figure}

As stated earlier, we use the Gaussian distribution's precision to model the channel noise. Results from our experiment are shown in Figure~\ref{fig:results}. The gamma posterior distribution is capable of tracking the actual noise precision (shown by the mean and one standard deviation in the top part of the figure). The BER shown in the bottom part of the figure is a moving average consisting of $10 000$ packets centred around the current point in time; this helps to reveal the underlying trend.
We use Dirichlet smoothing to avoid reporting BER values in the range $0 < \text{BER} < 9\times10^{-7}$, which is smaller than the possible BER. The BER shown in our results is calculated using: $ \text{BER} = \frac{1}{10000}\sum \frac{(a + \text{bit errors in a packet})}{(K + a)}$, where $a=0.005$ and $K$ is the number of message bits.

We observe an improvement in the BER compared to a PGM using a fixed value (a precision of 8.76 or 9.42 dB SNR) of the average noise precision (shown in the bottom part of the figure).

The BER performance of our proposed PGM closely follows the BER performance of the PGM with perfect knowledge of the noise precision. This is due to the small difference between the actual noise precision and the estimated noise precision of our proposed PGM.

In some experiments, we observed instances where the estimated precision was much lower than the actual precision. These were instances where the LDPC code failed to fix all bit errors, which the PGM then interpreted as worse than actual channel conditions. Another observation is that the gamma posterior has a wider standard deviation at higher precision means, and a narrower standard deviation at lower precision means. This is due to the characteristics of the gamma distribution. We view this as a beneficial artifact, since the PGM makes provision for more rapid changes in the channel conditions at lower channel noise. (This can also be regarded as a "sceptical" type of behaviour.) When channel conditions are bad, the PGM's channel estimation more confidently reports that they are bad.

A summary of the overall results appears in Table~\ref{tab:results}. The average BER of our PGM is slightly higher than the PGM with perfect knowledge of the channel, and outperforms the PGM with fixed value of the channel (at approximately 1.5 times better BER on average). Our PGM requires approximately 1.02 times fewer message passing iterations on average compared to the PGM with a fixed value of the channel, and the same number of iterations compared to the PGM with perfect knowledge of the channel.

\begin{table}[htb]
\centering
\caption{A summary of the overall performance comparison. Note that the expected behaviour of our PGM is similar to the PGM with perfect knowledge of the channel noise, and outperforms the PGM with fixed value of the average channel noise in terms of the BER.}
% \vspace{5mm}
\begin{tabular}[t]{ lcc }
& BER (mean): & Iterations (mean):  \\
\hline\hline
True precision & 0.003034 & 2.49 \\
\hline
Fixed precision & 0.005013 & 2.54 \\
\hline
Estimated precision & 0.003336 & 2.49 \\
\hline\hline
\end{tabular}
\label{tab:results}
\end{table}

Note that the averages presented in this table are more influenced by the high SNR instances where bit errors are infrequent. Figure~\ref{fig:results} illustrates the performance advantage of our proposed PGM more clearly where bit errors occur more frequently at lower SNRs.

% 0.003121750674947525 2.364618363912782
% 0.004763163150271192 1.864558719394113
% 0.0034896546329577847 2.4531409818756806

% 0.0030342105967243733 2.4966028005005065
% 0.005013693102200248 2.549041753701133
% 0.0033363786860185637 2.4924023035474527

\section{Supplementary investigation}
\label{sec:supplementary-investigation}
In some instances, the BER performance of our proposed PGM (estimating the SNR) is better than the PGM with knowledge of the actual SNR. We noted this happening when the estimated SNR is lower than the actual SNR, which is counter-intuitive.

A study presented in~\cite{mackay2003performance} investigates the effect of channel noise mismatch in LDPC codes. It found that the LDPC code's performance degrades more quickly when the assumed SNR is overestimated, but is less sensitive to degradation when underestimated up to around 1 dB below the actual SNR. What is also interesting is that the optimal BER is at an assumed SNR lower than the actual SNR.

We reproduce the same experiment to establish whether this behaviour is similar for a cluster graph representation of LDPC codes. Our channel noise mismatch experiment is presented as a BER vs model SNR curve where the model SNR is the assumed channel noise while the actual channel noise is fixed at 1.22 dB (similar to~\cite{mackay2003performance}). The purpose of the experiment is to (1) understand the impact on BER performance when the actual channel noise is under- and over-estimated by the model, and (2) determine where the optimal BER is.

\begin{figure}[!h]
     \centering
     \includegraphics[scale=0.4]{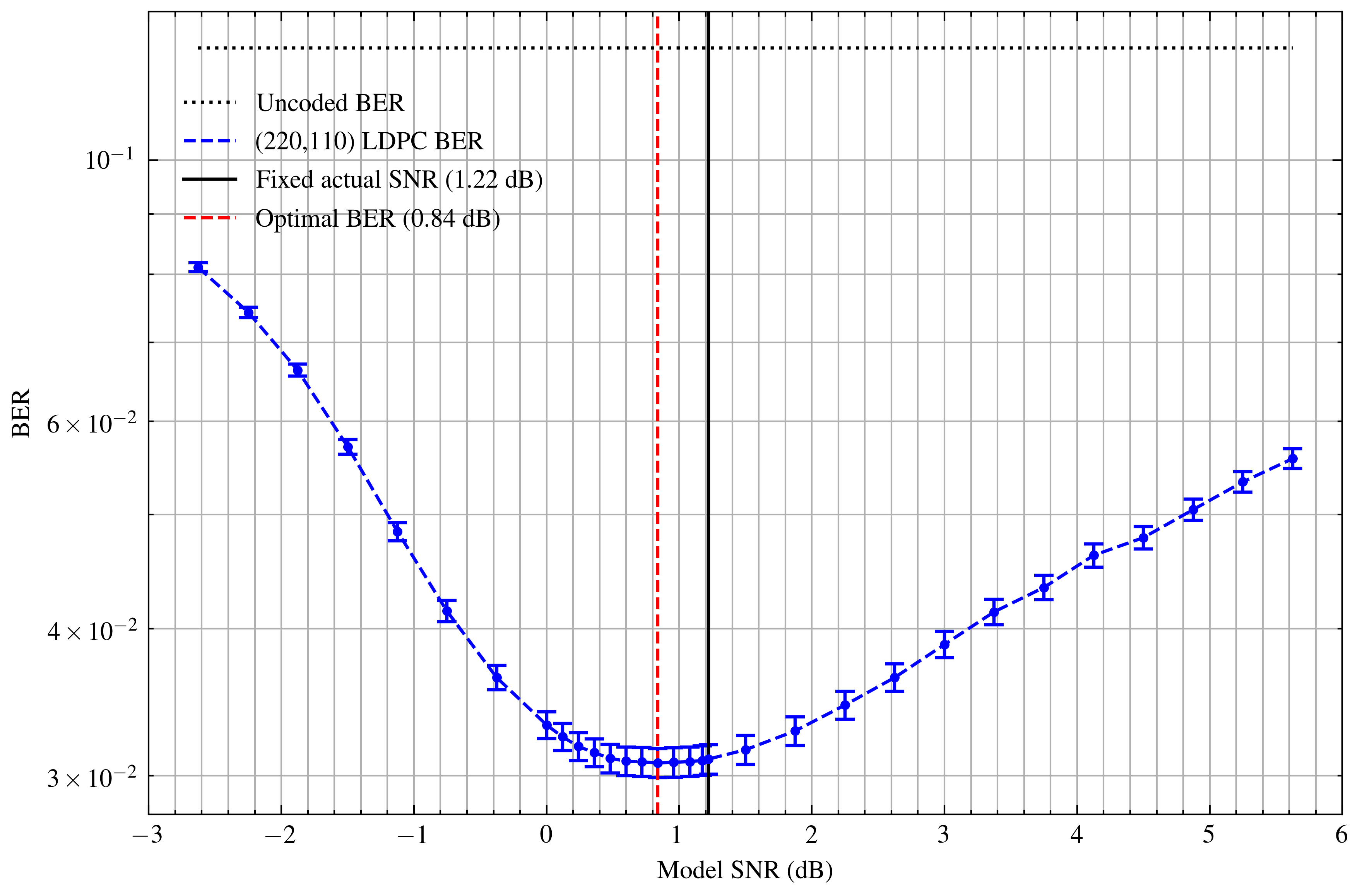}
     \caption{The BER performance of a cluster graph LDPC code with the actual SNR fixed at 1.22 dB when the model SNR is varied. The error bars reflect the 95\% confidence intervals.}
     \label{fig:mismatch}
\end{figure}

A random bit message is encoded and Gaussian noise is added using a fixed precision of 1.32 (equivalent to a SNR of 1.22 dB) to each bit, which repeats for 10k packets per model SNR value. The same received bit values are used for each model SNR value to ensure a like-for-like comparison. We set the maximum number of message passing iterations to 20. The results of the channel noise mismatch experiment is shown in Figure~\ref{fig:mismatch}.

This result is similar to~\cite{mackay2003performance} and suggests that the LDPC code's optimal BER is not necessarily at the point where the model SNR is equal to the actual SNR, but somewhere slightly below the actual SNR (in this case at 0.84 dB).

A conservative estimation of the channel noise seems to be beneficial, since the LDPC decoder is more forgiving in the region below the actual noise. Using the drive-test data we run the same experiment as described in Section~\ref{sec:non-stationary-experiment}. However, after each packet is decoded we adjust the estimated precision value to be -0.1 dB below the posterior gamma mean value. We found that an adjustment larger than this yielded performance degradation. The result of this adjustment is shown in Table~\ref{tab:results2}.

\begin{table}[htb]
\centering
\caption{A summary of the overall performance comparison between the PGM that estimates the channel noise and the PGM that more conservatively estimates the channel noise. Note that the expected behaviour of the more conservative approach performs better than our initial PGM only in terms of the BER.}
% \vspace{5mm}
\begin{tabular}[t]{ lcc }
& BER (mean): & Iterations (mean):  \\
\hline\hline
Estimated precision & 0.003336 & 2.49 \\
\hline
Estimated precision (-0.1 dB) & 0.003329 & 2.49 \\
\hline\hline
\end{tabular}
\label{tab:results2}
\end{table}
% 0.0033291567368103135 2.4975723411869515
A slight improvement in the BER can be seen, however, there is a trade-off between the number of iterations and BER performance. The decoder requires slightly more iterations on average when the adjustment to the posterior gamma is made.

We emphasise that the purpose of the supplementary investigation is not to propose that an adjustment be made to the estimated SNR in practice, but rather to show the counter-intuitive behaviour of the LDPC code that we found interesting. The benefit of adjusting the estimated SNR seems minimal.

\section{Conclusion and future work}
\label{sec:conclusion}
This paper contributes a simple sequential Bayesian learning method for tracking non-stationary channel noise using LDPC codes. We demonstrated the idea by employing a more general probabilistic framework called cluster graphs, and evaluated our proposed model using real-world data. The results show that the performance of our proposed model is nearly indistinguishable from a model with perfect knowledge of the channel noise.

Apart from the performance advantages shown in our results, the approach embeds well within an LDPC decoder and does not require stand-alone machine learning techniques, pre-training, or management of large data sets. It is capable of learning the channel noise accurately on-the-fly while decoding LDPC packets without compromise.

The implications of this method with respect to the communication system are that (1) the scheduling and use of pilot signals dedicated to channel noise estimation may become redundant, (2) LDPC codes coupled with our proposed channel-noise estimator can be used to estimate non-stationary channel noise on-the-fly with no need for scheduling, and (3) channel-noise information is now inherent in the LDPC code and provides performance advantages to the code.

Our current approach relies on parameterising a time constant, which is a prior assumption about the speed at which channel noise will vary. Future work will focus on applying real-time Bayesian model selection, allowing the LDPC decoder to choose intelligently between multiple assumptions (or combine multiple assumptions) about the speed at which the channel varies. Such work relates to Bayesian change-point detection, which accounts for changes in the underlying data-generating process while estimating parameters. While this study focuses on channel-noise estimation alone, the methodology could also be expanded to estimate other channel properties such as phase and complex channel gain, which will depend on the modulation scheme and channel model discussed in Section~\ref{sec:ldpc-with-snr-estimation}.

\section{Acknowledgements}
The authors would like to thank Bonnie Xaba for providing the drive-test data. She is a senior specialist in network benchmarking and E2E network performance quality at Vodacom Group Limited in South Africa.

\bibliographystyle{IEEEtran}
\bibliography{bibliography}  
% \bibliography{bibliography}%

\end{document}